\documentclass[12pt]{article}

\usepackage{scicite}
\usepackage[pdftex]{graphicx}
\usepackage{pgf}

\usepackage{times}

\newenvironment{sciabstract}{%
\begin{quote} \bf}
{\end{quote}}

\newcounter{lastnote}

\title{Engineering photonic density of states using metamaterials}

\author
{Zubin Jacob$^{1*}$, Ji-Young Kim$^{1*}$, Gururaj V. Naik$^1$,\\
Alexandra Boltasseva$^{1,2,3}$, Evgenii E. Narimanov$^1$, Vladimir M. Shalaev$^1$\\
\normalsize{$^{1}$Birck Nanotechnology Center, School of Electrical and Computer engineering,}\\ 
\normalsize{Purdue University, West Lafayette, IN 47907, U.S.A.,}\\
\normalsize{$^{2}$DTU-Fotonik Technical University of Denmark,}\\
\normalsize{ DTU Building 343, DK-2800 Kongens Lyngby, Denmark} \\
\normalsize{$^{3}$ Erlangen Graduate School in Advanced Optical Technologies (SAOT),}\\ 
\normalsize{Friedrich-Alexander-Universität Erlangen-Nürnberg, 91052 Erlangen, Germany}\\
\normalsize{$^\ast$These authors contributed equally to this work}
}

\date{}

\begin{document} 


\baselineskip24pt


\maketitle


\begin{sciabstract}
The photonic density of states (PDOS), like its' electronic counterpart, is one of the key physical quantities governing
a variety of phenomena and hence PDOS manipulation is the route to new photonic devices.
The PDOS is conventionally altered by exploiting the resonance within a device such as a microcavity  or a bandgap structure like a photonic crystal. Here we show that  nanostructured  metamaterials with hyperbolic dispersion can dramatically enhance the photonic density of states paving the way for metamaterial based PDOS engineering.
\end{sciabstract}

\clearpage

It was recently predicted that an appropriately designed metamaterial can be used to enhance the PDOS \cite{prediction} through a  nanopatterning route which presents a paradigm shift to PDOS alteration and can give rise to broadband, ultralow mode volume photonic devices, as opposed to conventional  resonant \cite{yablonovitch,spg1,spg2}, diffraction-limited approaches \cite{hughes,phc_purcell}. A simple probe for the local density of states is the spontaneous emission  of light in the vicinity of such a metamaterial \cite{barnes1}. In this work, we demonstrate that dye molecules in the near field of a hyperbolic metamaterial spontaneously excite unique electromagnetic states in the metamaterial medium responsible for the enhanced PDOS leading to a decrease in the lifetime of emitters.

The photonic density of states is obtained via a mode counting procedure in k-space from the dispersion relation $\omega(k)$ and can be simply related to the volume of the shell enclosed between two spheres at $\omega(k)$ and $\omega(k) + d\omega$ [Fig. 1(a)]. Here we show that nanostructuring a medium can radically alter the dispersion relation and hence allows the manipulation of the PDOS. Nanostructured metamaterials can have an extreme anisotropic electromagnetic response with $\epsilon_{x}=\epsilon_{y} < 0$ and $\epsilon_{z} > 0$ so that the extraordinary waves in this medium obey a hyperbolic dispersion relation \cite{indefinite,hyperlens,nader,hoffman,noginov}. While high wavevector modes simply decay away in vacuum, a hyperbolic metamaterial (HMM) allows bulk propagating waves with unbounded wavevectors [Fig. 1]. These spatial
modes with large wavevectors have been studied with regards to subwavelength imaging \cite{hyperlens_science,smolyaninov,shalaev} and subdiffraction
mode confinement \cite{viktor2}. The high wavevector spatial modes not present
in vacuum are characteristic of hyperbolic metamaterials (HMMs) and contribute to the photonic density of states causing a divergence in the low loss effective medium
limit \cite{pdos_original}. The remarkable property which sets it apart from other photonic systems with an enhanced DOS is the
bandwidth. Hyperbolic dispersion can be achieved in a large bandwidth \cite{hoffman,noginov} which directly implies that the
PDOS would in fact, diverge in this entire band of frequencies leading to a host of new physical effects.

The photonic density of states for a hyperbolic metamaterial has contributions from the large wavevector HMM states
up to the wavevector $k_{max}>>k_0$ and is given by \cite{pdos_original}
\begin{equation}
\rho(\omega) \sim k_{max}^{3}
\end{equation}
This presents the possibility of controlling the photonic density of states using nanostructured hyperbolic metamaterials in a broad bandwidth. In fact, in the effective medium limit the PDOS actually diverges
\cite{pdos_original}. Furthermore, since HMM states comprise of
high spatial wavevector modes, it can lead to PDOS manipulation in nanoscale mode volume metamaterial devices unlike a microcavity.


Among many effects which show a pronounced dependence on the photonic density of states is spontaneous emission \cite{purcell}. Fermi's
golden rule shows that a high PDOS immediately translates to a larger number of radiative decay channels available for an excited atom ensuring enhanced spontaneous emission into the particular mode of interest. We therefore use spontaneous emission to experimentally probe the HMM states and the enhanced density of states
of a hyperbolic metamaterial. While a rigorous quantum electrodynamic treatement allows the incorporation of
losses and material dispersion for metamaterial based radiative decay engineering \cite{hughes_niw}, the semiclassical principle \cite{ford} can be used in the weak coupling
limit \cite{hughes_niw} to clearly isolate the role of HMM states in the lifetime decrease of emitters placed in the vicinity of a hyperbolic metamaterial.

The lifetime of a dipole in the vicinity of nanostructures can be written in general as a sum of the lifetime contributions
due to various decay channels consisting of propagating waves in vacuum ($\Gamma^{vac}$), resonant routes such as plasmons or slow waveguide modes ($\Gamma^{res}$) and  a final route arising due to the evanescent waves with large spatial wavevectors ($\Gamma^{high-k}$).
\begin{equation}
\Gamma^{tot} = \Gamma^{vac} + \Gamma^{res} + \Gamma^{high-k}
\end{equation}
While a vast body of literature over many years has been devoted to the first two forms of radiative decay engineering \cite{wittke,kleppner,hughes_niw,ford,drexhage,barnes2,barnes3,rde}, the $\Gamma^{high-k}$ has been conventionally
associated with the phenomenon of quenching and non-radiative lifetime decrease occuring in the near field of any lossy object \cite{barnes1,ford}. Quite generally, the semiclassical principle \cite{ford} which relates the power carried by each of these routes to the respective decay rate contribution helps to identify the decay mechanism
in the near field for $d << \lambda$
\begin{equation}
\Gamma^{high-k} \sim \frac{Im(\epsilon)}{d^{3}}
\end{equation}
where $\epsilon$ is the dielectric constant of the structure placed at a distance $d<<\lambda$ from the emitting dipole. In the absence of losses, as in the case of a dielectric, the evanescent wave continuum route vanishes and $\Gamma^{high-k} \to 0$. For the case of metals, this high-k continuum is completely absorbed causing a non-radiative increase in the decay rate of the dipole; the effect known as quenching. In stark contrast to this, for a hyperbolic metamaterial, even in the limit of zero losses the contribution to the decay rate due to evanescent waves emanating from the dipole is
\begin{equation}
\Gamma^{high-k} \sim \frac{1}{d^{3}}\frac{2\sqrt{\epsilon_{z}|\epsilon_{x}|}}{1+\epsilon_{z}|\epsilon_{x}|}
\end{equation}
This helps to identify a completely different route which arises due to the excitation of HMM states within the metamaterial (see supplementary information). These are propagating waves with large spatial wavevectors (high-k). The presence of losses leads to a finite propagation length for these HMM states.  Note that the decay rate is enhanced radiatively in the near field of the HMM and the photons are emitted into the metamaterial.

The role of the HMM states is depicted in fig. (2) which shows the power emitted by a dipole as it is brought closer to the HMM. There is a high spatial wavevector continuum of states within the metmaterial which are excited by the dipole. These HMM states contribute to the lifetime
ultimately leading to very high decay rate as in Eq. (4) in the near field of the metamaterial. For comparison, a plot of the power emitted
by the dipole in the vicinity of a dielectric is shown. There is a sharp cut off that corresponds to the refractive index of the medium and the
high wavevector decay route is missing as expected from Eq. (3).

We now study the spontaneous excitation of HMM states using dye molecules. HMMs have been demonstrated with bulk properties and low loss at various wavelengths of interest. This makes them among the most promsing metamaterials for device applications due to their relative ease of fabrication \cite{hyperlens_science,hoffman,noginov}. Our HMM is designed (Fig. 3) to operate at the wavelength of emission of the chosen dye, Rhodamine 800.  Sixteen alternating subwavelength layers of alumina and gold (thickness 19 nm) are deposited using an electron beam evaporation technique with careful control of the individual layer thicknesses. Effective medium theory predicts that the dielectric permittivity of this metamaterial structure, in the direction parallel and perpendicular to the layers are of opposite signs achieving hyperbolic dispersion in a broad bandwidth. To confirm the effective medium calculations and hyperbolic dispersion at the wavelength of interest, the effective dielectric tensor is extracted from the reflectivity data for  s and p polarized  incident light (Fig. 3). The extracted value of the permittivity is $\epsilon_x=\epsilon_y=-8.2+0.9i$ and $\epsilon_z=12.5+0.5i$. The blue and red curves show the calculated reflectivities using the extracted value of the permittivity (see supplementary information).

Emission characteristics are studied using the Rhodamine 800 ($\lambda_{peak}  = 715 nm$) dye in an epoxy matrix ($n=1.58$) placed over the metamaterial. The thickness of the dye layer is 21 nm ascertained by AFM. A spacer layer provides isolation of the dye molecules from the
metal avoiding non-radiative transfer of energy and also ensures that the effective medium description is valid.  The lifetime measurements are carried out using flouorescence lifetime imaging (FLIM) system (MicroTime 200 Picoquant). The dye is excited using a pump laser ($\lambda_{pump}= 635 nm$) with 88 ps pulses.

The dye molecules (dispersed in 21 nm epoxy layer) placed on  metamaterial show a distinct decrease in lifetime as compared to the dielectric (Fig. 4(a),(b)). This is related to the fact that  the effective hyperbolic metamaterial allows spontaneous emission into surface plasmons and also high-k HMM states. Lifetime simulations of molecules on the metamaterial using a semiclassical approach based on a many dipole model give a factor of 1.7 decrease as compared to the dielectric (see supplementary information).  This is in agreement with the experimentally measured lifetimes of 2 ns for the dielectric and 1.1 ns for the HMM with a spacer of 21 nm. Similar results are obtained for a spacer layer of 29 nm as well.

The upper bound on the enhancement in the photonic density of states is caused by the metamaterial patterning scale. The effective homogenization theory for a metamaterial is valid only when the local wavelength of light is greater than the characteristic
metamaterial patterning scale $a$. For a metamaterial with the anisotropic response
$\epsilon_{x}=\epsilon_{y} < 0$ and $\epsilon_{z} > 0$, hyperbolic dispersion is achieved for wavevectors only up to an upper cut off
$k_{max} \sim 2\pi/a$. The simulations are performed taking into account an effective medium description of the hyperbolic
metamaterial. Another limiting factor is the loss inherent in current metamaterial technologies but recent developments in gain compensated metamaterials promise to usher in practical devices by utilizing active media to significantly enhance device performance.

Since the hyperbolic metamaterial necessarily requires a metal for achieving the extreme anisotropic electromagnetic response we systematically isolate the role of plasmons and quenching present due to losses. For comparison, we perform lifetime measurements on a gold substrate (305 nm thickness, same as the hyperbolic metamaterial) with dye molecules on top. The HMM sample shows a marked decrease in lifetime
as compared to the metal (1.6 ns) substrate as well (Fig. 4(a)). In the near field, the lifetime is dominated by the plasmonic states and the continuum of high-k wavevector states. The semiclassical model shows that the power carried (and hence the contribution to lifetime decrease) by plasmons on the metal or metamaterial are comparable. It also confirms that the excessive lifetime decrease for the metamaterial compared to the metal is caused by the continuum of high wavevector spatial modes emitted by the dye in the close vicinity of the metamaterial which couple to the HMM states. These high wavevector spatial modes cannot propagate in vacuum or dielectric and in the case of metal, they are simply absorbed, giving rise to non radiative decrease in the lifetime (quenching). Our simulations show that the imaginary part of the permittivities which have been ascertained with high accuracy (and confirmed using ellipsometry) do not cause any additional quenching as compared to the metal (see supplementary information).  The hyperbolic dispersion in the metamaterial allows propagating  modes with large wavevectors which is the cause for a radiative decrease in the lifetime of the dye.

To summarize, we have made the first step in a metamaterial based route to engineer the photonic density of states.  We have probed the available local density of states in the near field of the metamaterial using spontaneous emission of dyes. The decrease in the lifetime of emitters demonstrates the spontaneous excitation of unique electromagnetic states in the metamaterial. For a low loss metamaterial with deep subwavelength nanopatterning these unique electromagnetic states can far exceed the vacuum density of states dominating all processes that are governed by the PDOS. The use of gain media to compensate losses in metallic environments has led to many significant developments like the SPASER \cite{spaser,oulton} and such approaches will lead to a new methodology of PDOS engineering based on hyperbolic metamaterials.

\clearpage

\begin{figure}
\scalebox{0.35}{\includegraphics{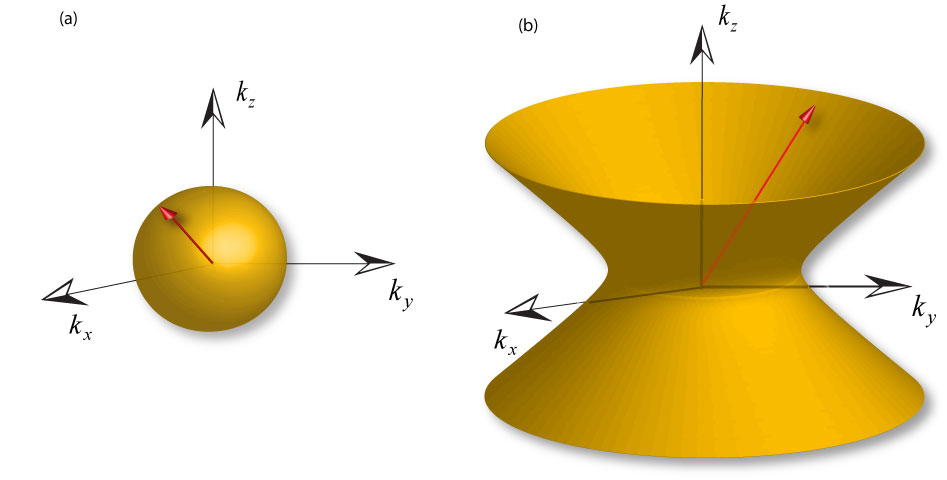}}
\caption{\label{fig1} (a) Dispersion relation for an isotropic medium. The red arrow denotes an allowed wavevector. The photonic density of states is related to the volume of a shell bounded between two such spheres corresponding to two frequencies (b) Hyperbolic dispersion relation allowing unbounded values of the wavevector (red arrow) due to which the photonic density of states diverges.}
\end{figure}

\clearpage

\begin{figure}
\scalebox{0.5}{\includegraphics{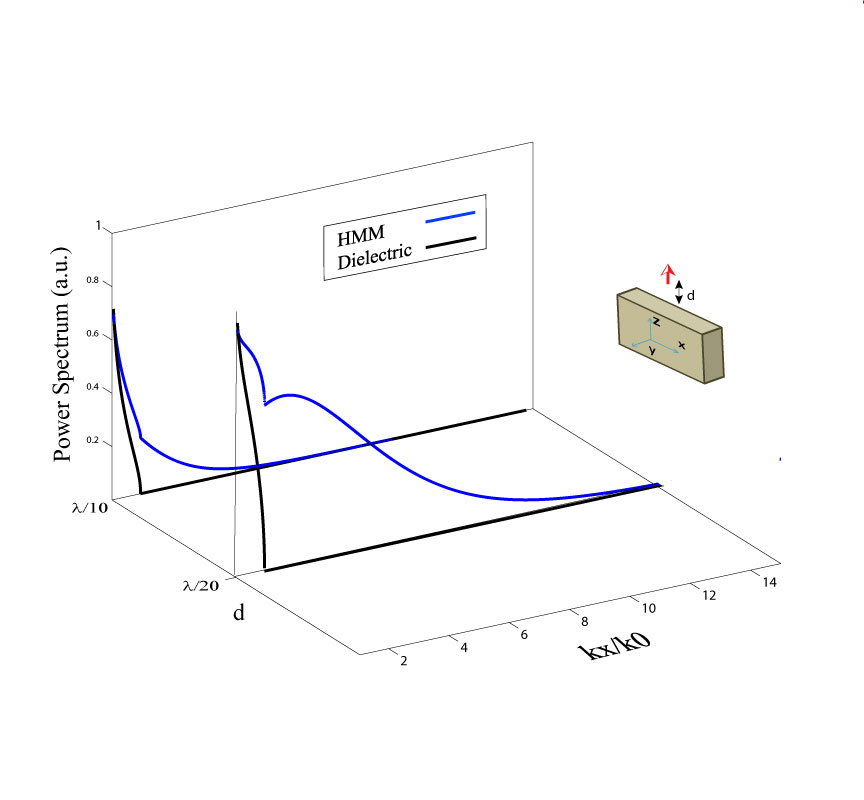}}
\caption{\label{fig2}  The distribution of power emitted by a dipole (red arrow) as it is brought closer to a slab of dielectric (black curve) or HMM (blue curve). In the close vicinity of the metamaterial, power is transferred via evanescent waves in vacuum to the high wavevector HMM states which causes the lifetime to decrease as compared to the dielectric}
\end{figure}

\clearpage

\begin{figure}
\scalebox{0.5}{\includegraphics{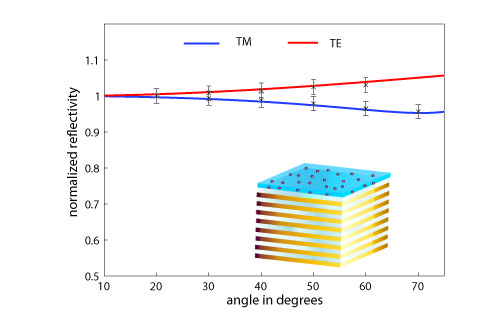}}
\caption{\label{fig3}  Reflection measurements on the metamaterial to characterise the desired anisotropic electromagnetic response. Normalized reflectivity is R($\theta$)/R(20$^{0}$). Schematic shows the fabricated hyperbolic metamaterial consisting of alternating subwavelength layers of metal (gold) and dielectric (alumina). There is a spacer layer on top separating the metamaterial from a dye dispersed in an epoxy medium. }
\end{figure}

\clearpage

\begin{figure}
\scalebox{0.5}{\includegraphics{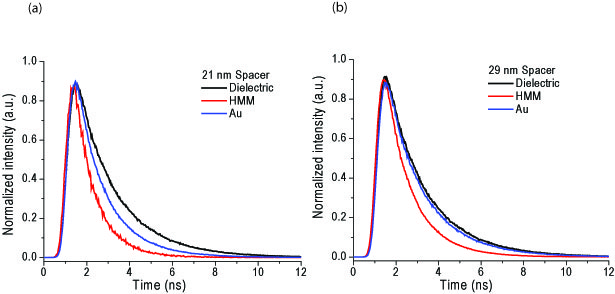}}
\caption{\label{fig4} (a) and (b) Comparison of the spontaneous emission lifetime of dye molecules on top of a dielectric (black), metal(blue) and metamaterial (red).  Systematic isolation of the effects due to quenching and plasmons can be achieved by
comparing the HMM to a metal slab (blue curve). The decrease in lifetime on top of the metamaterial as compared to the metal is due to the spontaneous excitation of HMM states by dye molecules in the near field of the HMM.}
\end{figure}



\end{document}